\documentclass[square]{ws-procs975x65}

\begin{document}

\title{Virial masses of galaxy clusters in the post-Newtonian limit}

\author{Mahmood Roshan}

\address{School of Physics, Institute for Research in Fundamental Sciences (IPM),\\
Tehran, Iran\\
$^*$E-mail: roshan@ipm.ir}

\begin{abstract}
We estimate virial masses of galaxy clusters using the parametrized post-Newtonian (PPN) virial theorem. Also, we show explicitly that post-Newtonian corrections can not address the mass discrepancy in the galaxy clusters.
\end{abstract}

\keywords{Virial mass; Post-Newtonian virial theorem.}

\section{Introduction}

It is known that the intracluster medium (ICM), hot X-ray emitting gas, contains most of the baryonic material in a galaxy clusters. Since, the mean free paths of electrons and protons in the ICM are very smaller than the characteristic radius of the galaxy cluster, then it is reasonable to treat the ICM as a fluid. Thus, for analyzing the ICM we can use the parametrized post-Newtonian (PPN) virial theorem \cite{roshan}. This PPN virial theorem has been derived for a hydrodynamic system\cite{roshan}. On the other hand, using the ideal gas law the pressure and the matter density are related as $p(r)=\sigma_r^2 \rho(r)$,
where $
\sigma_r^2=\frac{k_B T(r)}{\mu m_p}$
is the velocity dispersion in the radial direction, $k_B$ is Boltzmann's constant, $\mu=0.609$ is the mean atomic weight, $m_p$ is the proton mass and $T(r)$ is the temperature profile. Also, the internal energy per unit mass of the ideal gas is given by
$
\Pi(r)=\frac{\sigma_r^2}{\Gamma-1}
$
where $\Gamma$ is the adiabatic index. If we assume that ICM is a isothermal spherical configuration, then $T(r)$, $p/\rho$ and $\Pi$ are constants. Now, let us simplify the scalar PPN virial theorem as follows
\begin{equation}
\begin{split}
&\int \rho v^2 d^3x-\frac{1}{2}\int \rho U d^3x+3\int p d^3x+\int \rho v^2 \left(\Pi+v^2+\frac{p}{\rho}\right)d^3x+4\gamma\int \rho v^2 U d^3x\\&+(6\gamma-4)\int pU d^3x-\int \rho U \Pi d^3x-(3\gamma-2\beta+1)\int \rho U^2 d^3x-\frac{4\gamma+3+\alpha_1-\alpha_2}{4} \mathfrak{u}\\&-\frac{1+\alpha_2}{4} Z+\xi M=0
\end{split}
\label{vir3}
\end{equation}
where $v$ is the velocity, $\alpha_1$, $\alpha_2$, $\gamma$, $\xi$ and $\beta$ are PPN parameters, $U$ is the Newtonian gravitational potential and the potentials $Z$, $M$ and $\mathfrak{u}$ have been introduced in \cite{roshan,chandra}. In order to make an estimation of the virial mass, it is convenient to introduce the virial mass $M_V$ and the virial radius $R_V$ as follows
\begin{equation}
\frac{1}{2}\int \rho U d^3x=\frac{3}{5} M_{gas}\left(\frac{M_V}{R_V}\right)
\label{vir4}
\end{equation}
where $M_{gas}$ is the total mass of the hot gas. We may also write
\begin{equation}
\int \rho U^2 d^3 x=\frac{17}{15}M_{gas}\left(\frac{M_V}{R_V}\right)^2
\label{vir44}
\end{equation}
After elementary calculations the integrals of equation \eqref{vir3} can be rewritten as
\begin{equation}
\begin{split}
&\int \rho v^2 d^3x\simeq M_{gas} \overline{v^2},~~~~~~~~\int pd^3x=M_{gas} \sigma_r^2,~~~~Z \sim \mathfrak{u}\simeq M_{gas}\left(\frac{M_V}{R_V}\right) \overline{v}^2\\&
\int \rho v^2 \left(\Pi+v^2+\frac{p}{\rho}\right)d^3x\simeq M_{gas} \overline{v^4}+\frac{\Gamma}{\Gamma-1}\sigma_r^2\overline{v^2}M_{gas}\\&
\frac{1}{2}\int\rho v^2 U d^3x\simeq  \frac{3}{5} M_{gas}\left(\frac{M_V}{R_V}\right) \overline{v^2},~~~~~~~
\frac{1}{2}\int U p d^3x=\frac{3}{5} M_{gas}\left(\frac{M_V}{R_V}\right) \sigma_r^2\\&
\frac{1}{2}\int \rho U \Pi d^3x=\frac{3}{5} M_{gas}\left(\frac{M_V}{R_V}\right) \frac{\sigma_r^2}{\Gamma-1},~~~~~M\simeq -2 M_{gas}\left(\frac{M_V}{R_V}\right)^2
\end{split}
\label{integrals}
\end{equation}
where $\overline{v^2}=3\sigma_r^2$, $\overline{v^4}=15\sigma_r^4$ and  $\overline{v}=\sqrt{\frac{8}{\pi}}\sigma_r$. By substituting equations \eqref{vir4}, \eqref{vir44} and \eqref{integrals} into equation \eqref{vir3}, and assuming that the ICM is composed of a monatomic gas ($\Gamma=5/3$), we get
\begin{equation}
\frac{M_V}{R_V}=\frac{30 \sigma_r ^2 (\alpha_1 +4 \gamma +4)+9 \pi  \left((9-36 \gamma ) \sigma_r ^2+1\right)\pm X}{2 \pi  (34 \beta -51 \gamma -30 \xi -17)}
\label{vir6}
\end{equation}
where
\begin{equation}
\begin{split}
&X=3 [\left (10 \sigma_r ^2 (\alpha_1 + 4 \gamma + 4) +
       3 \pi  \left ((9 - 36 \gamma ) \sigma_r ^2 +
          1 \right) \right)^2 -\\&
  ~~~~~~ 10 \pi ^2 \sigma_r ^2 \left (15 \sigma_r ^2 + 4 \right) (34 \beta - 51
          \gamma - 30 \xi - 17)]^{\frac{1}{2}}
\end{split}
\end{equation}
We choose the minus sign for $X$ because it recovers the Newtonian virial theorem in the Newtonian limit. On the other hand, in the galaxy clusters $\sigma_r^2\ll 1$,
thus we have $\frac{M_V}{R_V}\simeq10 \sigma_r ^2- n \sigma_r^4+\mathcal{O}(\sigma_r^6)$
%\begin{equation}
%\frac{M_V}{R_V}\simeq10 \sigma_r ^2- n \sigma_r^4+\mathcal{O}(\sigma_r^6)
%\label{vir7}
%\end{equation}
where
\begin{equation}
n=\frac{5}{18 \pi }( 120 (\alpha_1 +4 \gamma +4)+\pi  (-1360 \beta +744 \gamma +1200 \xi +869))
\label{n1}
\end{equation}

The term $n\sigma_r ^4$ is the post-Newtonian correction to the virial mass of galaxy clusters. Using the X-ray properties of 30 clusters \cite{reip}, we obtain $n\sim 10^6$ in order to get $M_V\sim M_{gas}$ . However, substituting the current values of PPN parameters \cite{will2} into \eqref{n1}, we get $n\sim 155.5$. It is obvious that the post-Newtonian correction is not large enough to solve the mass problem in the galaxy clusters.
\section*{Acknowledgments}
The author thanks the National Elite's Foundation of Iran for financial support.

\end{document}